\documentclass[aip,graphicx]{revtex4-1}

\usepackage[margin=1in]{geometry}
\usepackage{amsmath,amsthm,amsfonts,mathtools}
\usepackage{subcaption,caption}
\usepackage[final]{graphicx}
\usepackage{booktabs,tabularx,multirow}
\usepackage{microtype}
\usepackage{hyperref}
\usepackage{tikz}
\usepackage{amsthm}

\begin{document}

\title{Isolas of limit cycles and birhythmicity induced by cooperative feedback in a glycolysis model}



\author{Fangyuan Wang}
\email[]{freya{\_}fangyuanw@163.com}
\affiliation{Department of Mathematics, China Jiliang University, Hangzhou 310018, China}

\author{Lendert Gelens$^{*}$}
\affiliation{Laboratory of Dynamics in Biological Systems, Department of Cellular and Molecular Medicine, KU Leuven, Herestraat, 49, Leuven, Belgium}

\author{Yancong Xu$^{*}$}
\affiliation{Department of Mathematics, China Jiliang University, Hangzhou 310018, China}

\author{Libin Rong}
\email[]{libinrong@ufl.edu}
\affiliation{Department of Mathematics, University of Florida, Gainesville, 32611, FL, USA \\ 
*Corresponding authors. \\ Emails: lendert.gelens@kuleuven.be, Yancongx@cjlu.edu.cn}



\date{\today}

\begin{abstract}
We investigate how cooperative feedback shapes global oscillatory dynamics in a glycolysis model with product recycling and allosteric phosphofructokinase regulation.
Using bifurcation theory and numerical 
continuation, we analyze the stability of equilibria and characterize 
Hopf and generalized Hopf bifurcations, using the Hill exponent as an 
effective measure of cooperativity. We show that a codimension-2 
cusp-of-cycles point governs the creation and annihilation of detached 
branches of limit cycles (isolas) and, together with saddle-node 
bifurcations of limit cycles, organizes a regime map of six qualitatively 
distinct dynamical regions. In the birhythmic regime two stable oscillatory states coexist on connected branches; in the isola regime a stable oscillation exists on a fully disconnected branch, producing threshold-dependent onset of rhythmic activity.
 Time-domain simulations confirm coexistence of distinct rhythms 
and illustrate how the choice of initial condition determines which 
attractor is reached. Together, these results show how variations in 
cooperative feedback strength can generate isolated oscillatory modes and 
multistability in metabolic networks, highlighting isola dynamics as a 
general mechanism for rhythm selection and switching in nonlinear 
biological oscillators.
\end{abstract}

\maketitle

\section{Introduction}

Glycolysis is one of the most studied metabolic pathways in biology, and a 
classical example of nonlinear dynamics. Oscillatory behavior in glycolytic 
reactions was first observed experimentally by Duysens and Amesz 
\cite{duysens1957} and has since been widely studied in biochemical, physical, 
and mathematical contexts \cite{chance1964,ghosh1964,pye1966}. These 
oscillations arise from feedback loops involving the enzyme 
phosphofructokinase (PFK), whose allosteric regulation introduces strong 
nonlinearities into the reaction network.

A striking feature of glycolytic models is \emph{birhythmicity}: the 
coexistence of two stable oscillatory states with different amplitudes and 
periods under identical conditions \cite{pisarchik2014,biswas2023}. This 
arises from the interplay of product-activated feedback, substrate depletion, 
and open metabolic fluxes \cite{markus1984,ghosh1964}. Classical PFK models 
reproduce sustained oscillations but typically admit only a single periodic 
state \cite{goldbeter1972}, while extensions incorporating additional feedback 
or product recycling can produce nested stable limit cycles separated by 
unstable periodic orbits \cite{decroly1982,biswas2023,gustavsson2014,
jaiswal2024}. Recent work has further shown that birhythmic oscillators can 
undergo rate-induced transitions between coexisting rhythms under slowly 
varying parameters~\cite{kumar2025}. From a biological perspective, the 
ability to sustain multiple oscillatory modes may enable metabolic flexibility 
and adaptive responses to environmental changes.

A key control over this behavior is the degree of \emph{cooperativity} in 
allosteric feedback — how sharply a regulatory response switches as a function 
of substrate or product concentration. We model this via Hill-type 
nonlinearities, using the Hill exponent~$n$ as an effective measure of 
cooperativity (ultrasensitivity).
For PFK in yeast, Hill coefficients in the range 
$n \approx 2$--$4$ have been reported experimentally~\cite{gustavsson2014}, 
placing the cooperativity values studied here in a biologically relevant range. 
Previous work on related glycolysis models has shown that feedback strength 
can give rise to saddle-node bifurcations of limit cycles and codimension-2 
cusp structures that organize birhythmic regimes~\cite{biswas2017}. What has 
not been studied is how cooperativity can produce \emph{isolas} of limit 
cycles — isolated, disconnected branches of oscillatory solutions — in such 
systems.

An isola is a closed branch of periodic solutions bounded on both sides by 
saddle-node bifurcations, with no connection to the equilibrium through a 
Hopf bifurcation \cite{detroux2018,giri2021,sandstede2012,aougab2019}. Isolas 
have been found across many physical, chemical, and biological systems and are 
associated with multistability, hysteresis, and abrupt state switching 
\cite{karst2025,xu2025,xu2020,zhu2024,xu2025b,xu2025c}. They emerge naturally 
when S-shaped hysteresis curves interact: if two such curves are connected, 
they form a mushroom-shaped bifurcation diagram; if the connection is severed, 
an isolated branch appears \cite{ganapathisubramanian1984,giri2021}. As a 
parameter is varied, an isola can shrink to a point — the \emph{isola center} 
— and disappear \cite{yang2024,sandstede2012,xu2025c}.

Although both isolas and birhythmicity involve coexisting oscillatory states, 
they are geometrically and dynamically distinct. In birhythmicity, two stable 
limit cycles arise on branches connected to the equilibrium through Hopf 
bifurcations, and transitions between them require crossing an unstable 
separating orbit. In the isola regime, a stable oscillation exists on a fully 
disconnected branch with no Hopf connection: rhythmic activity can only be 
reached by a sufficiently large perturbation, producing a threshold-dependent, 
all-or-nothing onset of oscillations qualitatively unlike the smooth amplitude 
growth at a classical Hopf bifurcation. Systematically identifying which 
regime occurs under which parameter conditions has not previously been done 
for glycolytic models.

In this paper we show that varying cooperativity in a glycolysis model with 
allosteric feedback and product recycling systematically generates and destroys 
such isolas, providing a new mechanism for transitions between monorhythmic 
and birhythmic dynamics. We use bifurcation theory and numerical continuation 
to map the global structure of periodic solutions across parameter space, 
identifying six distinct dynamical regimes organized by a codimension-2 
cusp-of-cycles point as the key organizing center for these transitions. 
Time-domain simulations illustrate coexistence of rhythms and how the choice 
of initial condition determines which attractor is reached.

The paper is organized as follows. Section~\ref{sec:model} introduces the
glycolysis model, establishes existence and uniqueness of the positive
equilibrium, and outlines the local and global bifurcation tools used
throughout. Section~\ref{sec:global} presents the global bifurcation
structure and the dynamical regimes it defines. Section~\ref{sec:time}
examines time-domain dynamics and attractor coexistence.
Section~\ref{sec:conclusion} concludes the work and detailed derivations of the
Lyapunov coefficients and Hopf conditions are collected in the appendices.

\begin{figure}[htbp]
    \centering
    \includegraphics[width=0.9\columnwidth]{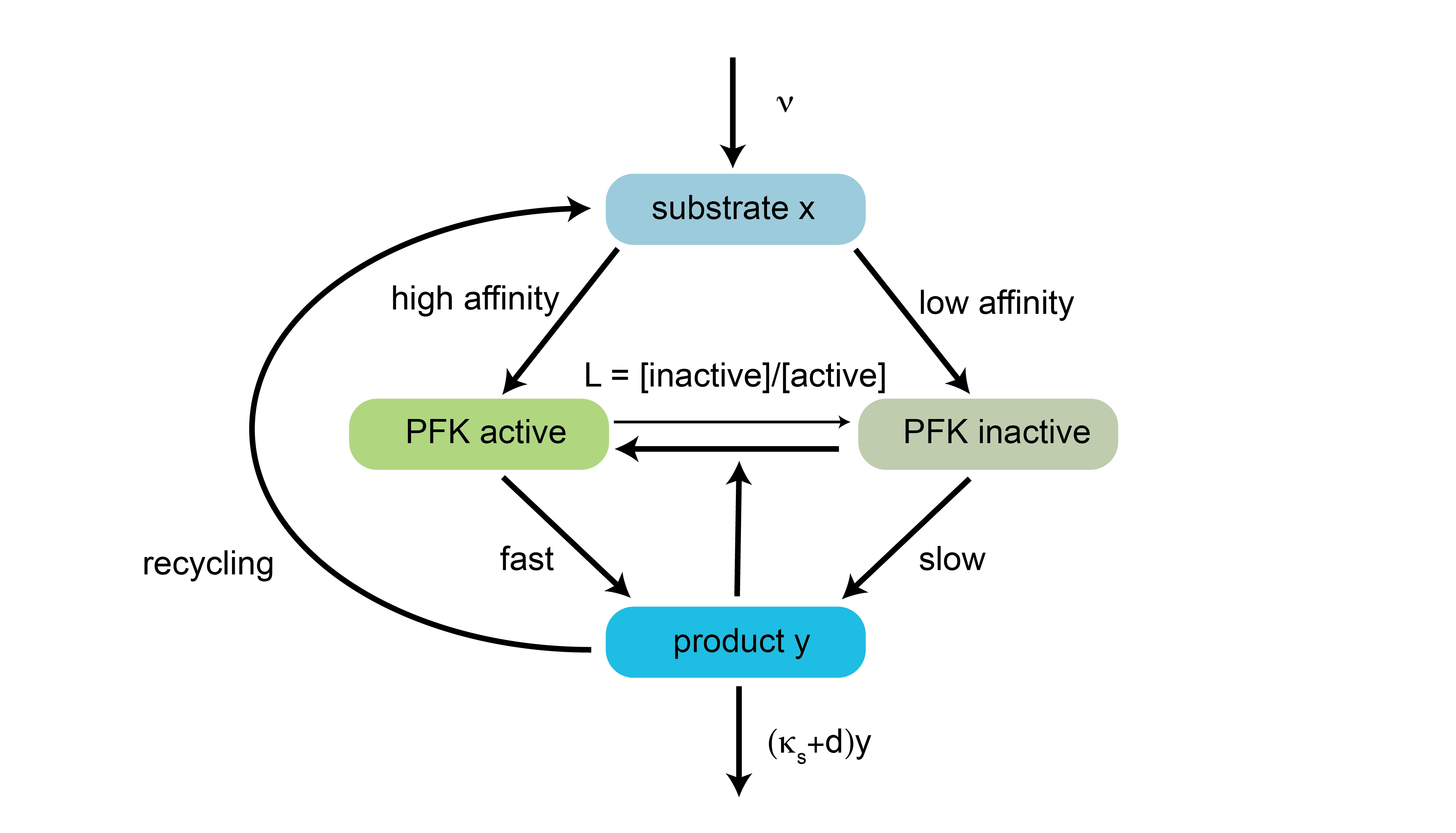}
    \caption{Schematic of the glycolysis model with allosteric feedback and 
    product recycling~\cite{kar2004}. Substrate~$x$ is converted to 
    product~$y$ via the PFK enzyme, which switches between an active state 
    (high affinity, fast catalysis) and an inactive state (low affinity, slow 
    catalysis), with $L=[\text{inactive}]/[\text{active}]$. Product is 
    recycled to substrate at rate $\sigma_i y^n/(\kappa^n+y^n)$ (solid red 
    arc) and removed at rate $(\kappa_s+d)y$. Elevated product activates PFK 
    by shifting the conformational equilibrium toward the active state (dashed 
    arrow), constituting the positive feedback loop responsible for oscillatory 
    dynamics.}
    \label{fig1}
\end{figure}

\section{Model equations and mathematical framework}
\label{sec:model}

We consider a modified Decroly--Goldbeter glycolysis model with allosteric
PFK regulation and product recycling~\cite{kar2004}. The system tracks two
normalized concentrations: substrate~$x$ and product~$y$, whose dynamics are
illustrated in Fig.~\ref{fig1}. Three processes govern the evolution of the
system: (i)~substrate inflow at rate~$\nu$; (ii)~conversion of substrate to
product through the allosterically regulated PFK enzyme; and (iii)~recycling
of product back to substrate, together with product removal at
rate~$(\kappa_s + d)y$.

PFK switches between an active and an inactive conformational state. The
ratio $L = [\text{inactive}]/[\text{active}] \gg 1$ in the absence of ligand,
so the enzyme is predominantly inactive at rest. The active form binds
substrate with higher affinity and converts it at a higher rate than the
inactive form. Product~$y$ acts as an allosteric activator: elevated~$y$
shifts the conformational equilibrium toward the active state, increasing the
conversion rate of~$x$ to~$y$. This constitutes a positive feedback loop.
The recycling term follows Hill-type kinetics with exponent~$n$, which
controls how sharply the recycling response switches with~$y$ and serves as
an effective measure of cooperative feedback.

The governing equations are
\begin{equation}
\begin{aligned}
\frac{dx}{dt} &= \nu - \frac{\sigma x(x+1)(y+1)^2}{L + (x+1)^2(y+1)^2}
+ \frac{\sigma_i y^n}{\kappa^n + y^n},\\[6pt]
\frac{dy}{dt} &= \frac{q\sigma x(x+1)(y+1)^2}{L + (x+1)^2(y+1)^2}
- \frac{q\sigma_i y^n}{\kappa^n + y^n} - (\kappa_s+d)\,y,
\end{aligned}
\label{eq:model}
\end{equation}
where all parameters are positive except $d$, which may be negative. The
parameter~$\sigma$ sets the maximum rate of the PFK reaction; $\nu$ is the
substrate inflow rate; $\sigma_i$ is the maximum recycling rate of product
back to substrate; and $\kappa_s$ is the basal product removal rate. Two
parameters are central to this study. The Hill exponent~$n$ controls the
effective cooperativity of the recycling feedback: larger~$n$ produces a
sharper, more switch-like response. The parameter~$d$ modulates the net
removal rate of product and thus the overall flux balance. Together they act
as complementary control knobs---$n$ for nonlinear amplification and $d$ for
flux balance---and their combined variation reshapes the global dynamical
structure of the system.

\subsection*{Equilibria and local stability}

Solutions of~\eqref{eq:model} remain non-negative and bounded for all
$t \geq 0$, confining the biologically meaningful dynamics to the first
quadrant. Equilibria are determined by the intersection of the nullclines
of~\eqref{eq:model}. Adding $1/q$ times the second equation to the first
eliminates the nonlinear flux terms and gives the explicit relation
$y^* = q\nu/(\kappa_s + d)$, which is positive whenever $\kappa_s + d > 0$.
Substituting $y^*$ reduces the $x$-nullcline to a quadratic $F(x^*) = 0$
whose coefficients are given in Appendix~\ref{app:equilibrium}. By
Descartes' rule of signs this quadratic admits at most one positive root,
and a unique positive equilibrium $E^* = (x^*, y^*)$ exists if and only if
\begin{equation}
  \sigma > \nu + \frac{\sigma_i q^n \nu^n}{\kappa^n(\kappa_s+d)^n + q^n\nu^n}.
  \label{eq:exist}
\end{equation}

Local stability of $E^*$ is governed by the Jacobian $J(E^*)$. A direct
computation (Appendix~\ref{app:hopf}) shows $\det(J(E^*)) > 0$ whenever
$\kappa_s + d > 0$, ruling out saddle equilibria and Bogdanov--Takens
bifurcations. Stability therefore reduces to the sign of
$\operatorname{tr}(J(E^*))$, and a Hopf bifurcation occurs when
\begin{equation}
  \operatorname{tr}(J(E^*)) = 0, \qquad \det(J(E^*)) > 0.
  \label{eq:hopf}
\end{equation}

\subsection*{Hopf criticality and Lyapunov coefficients}

To classify Hopf bifurcations and detect codimension-2 degeneracies, we
reduce the flow near $E^*$ to a planar normal form and compute Lyapunov
coefficients using the algorithm of Yu~\cite{Yu1998}. The sign of the first
Lyapunov coefficient $\sigma_1$ determines whether the bifurcation is
supercritical ($\sigma_1 < 0$, stable limit cycle born) or subcritical
($\sigma_1 > 0$, unstable limit cycle born). At a generalized Hopf (Bautin)
point $\sigma_1 = 0$; the sign of the second Lyapunov coefficient $\sigma_2$
then determines the local two-cycle configuration and the birth of a
saddle-node-of-cycles branch. Full expressions for $\sigma_1$ and $\sigma_2$
at $n = 2$, where the PFK flux simplifies to
\begin{equation}
\Phi(x,y)=\frac{x(1+x)(1+y)^2}{L+(1+x)^2(1+y)^2},
\end{equation}
are given in Appendix~\ref{app:lyapunov}. We treat $n$ as a bifurcation
parameter in the numerical continuation described below.

\subsection*{Global continuation strategy}

Numerical continuation with AUTO-07P~\cite{Doedel2007} tracks Hopf curves
and saddle-node-of-limit-cycle (SNL) curves in two-parameter planes,
starting from the Hopf points located analytically via~\eqref{eq:hopf}.
Generalized Hopf points on these curves seed the SNL branches, and the
termination of SNL branches at a codimension-2 cusp-of-cycles point (CPL)
signals the creation and annihilation of isolated branches (isolas) of
periodic solutions. Together, local invariants at $E^*$ and the
globally-continued bifurcation curves provide a complete picture of the
periodic-orbit families characterized in Section~\ref{sec:global}.

\begin{table}[h]
\centering
\caption{Baseline parameter values used throughout
Sections~\ref{sec:global} and~\ref{sec:time}. The Hill exponent~$n$ and
feedback parameter~$d$ are treated as bifurcation parameters and varied
throughout the analysis.}
\label{tab:params}
\begin{tabular}{lll}
\hline\hline
Symbol & Description & Value \\
\hline
$\nu$       & substrate inflow rate               & $0.255$ \\
$q$         & stoichiometric ratio                & $1.0$   \\
$\kappa_s$  & basal product removal rate          & $0.06$  \\
$L$         & conformational equilibrium ratio    & $3.6\times10^{6}$ \\
$\sigma$    & maximum PFK reaction rate           & $10.0$  \\
$\sigma_i$  & maximum recycling rate              & $1.3$   \\
$\kappa$    & recycling half-saturation constant  & $10.0$  \\
\hline\hline
\end{tabular}
\end{table}

\begin{figure}[htbp]
\hspace{-1.5cm}
    \centering
\includegraphics[width=0.9\columnwidth]{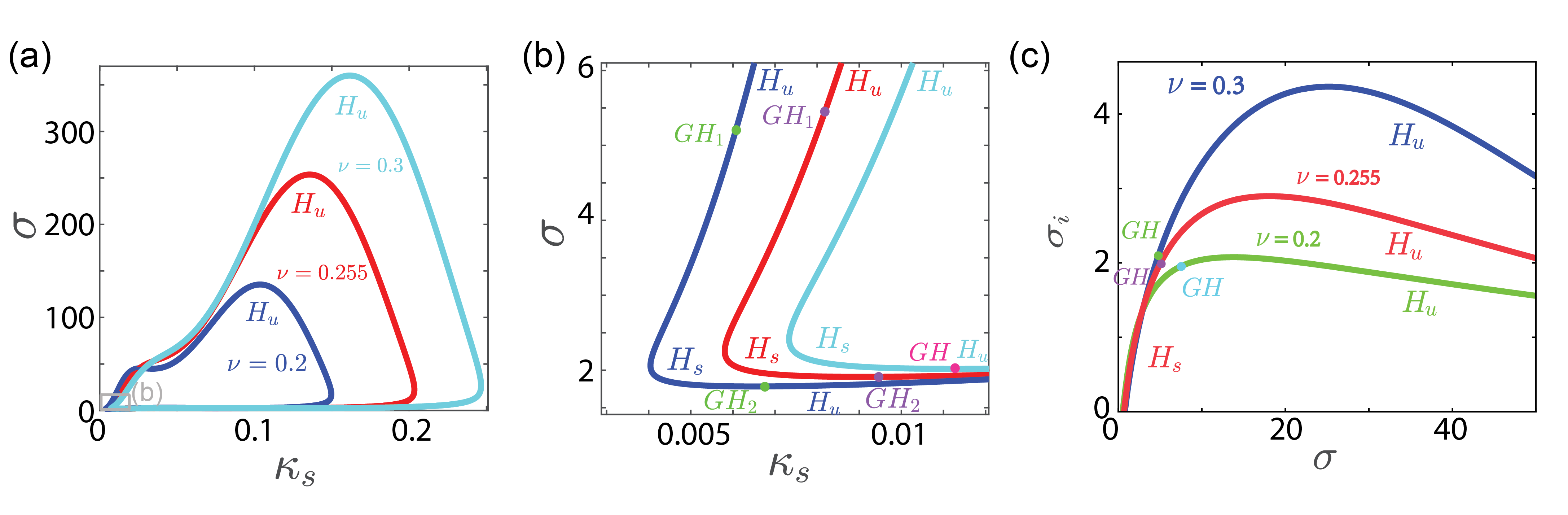}
    \caption{Two-parameter Hopf bifurcation diagrams. (a)~Hopf curves in the 
    $(\kappa_s,\sigma)$-plane for $\nu=0.2$, $0.255$, $0.3$. (b)~Zoom of 
    panel~(a). (c)~Hopf curves in the $(\sigma_i,\sigma)$-plane for 
    $\nu=0.2$, $0.255$, $0.3$. Generalized Hopf (Bautin) points mark changes 
    in Hopf criticality and the birth of saddle-node-of-cycles branches.}
    \label{fig:three_figures}
\end{figure}

\section{Global bifurcation structure and dynamical regimes}
\label{sec:global}

We now describe how the global organization of oscillatory solutions depends 
on cooperativity and flux balance. Throughout this section, numerical 
continuation is performed with AUTO-07P~\cite{Doedel2007}. Unless otherwise 
stated, parameter values are those listed in Table~\ref{tab:params}, taken 
from Ref.~\cite{biswas2017}.

\subsection{Hopf bifurcation curves and generalized Hopf points}

Figure~\ref{fig:three_figures} shows closed Hopf bifurcation curves in two 
parameter planes for $\nu = 0.2$, $0.255$, and $0.3$. In both planes the 
enclosed Hopf region shrinks as $\nu$ decreases. Along each curve the Hopf 
criticality changes sign at generalized Hopf (Bautin) points, which are 
listed in Table~\ref{tab:GH} and labeled $GH$ in the figure. At each such 
point a branch of saddle-node bifurcations of limit cycles is born. The 
number of generalized Hopf points depends on~$\nu$: two arise in the 
$(\kappa_s,\sigma)$-plane for $\nu=0.2$ and $\nu=0.255$, but only one for 
$\nu=0.3$, while a single generalized Hopf point appears in the 
$(\sigma_i,\sigma)$-plane for each value of~$\nu$.

\begin{table}[h]
\centering
\caption{Generalized Hopf (Bautin) point coordinates in the two parameter 
planes of Fig.~\ref{fig:three_figures}.}
\label{tab:GH}
\begin{tabular}{llll}
\hline\hline
Plane & $\nu$ & Point & Coordinates \\
\hline
\multirow{5}{*}{$(\kappa_s,\,\sigma)$}
  & \multirow{2}{*}{$0.2$}
      & $GH_1$ & $\kappa_s = 6.084\times10^{-3},\quad \sigma = 5.232$ \\
  &   & $GH_2$ & $\kappa_s = 6.721\times10^{-3},\quad \sigma = 1.784$ \\[3pt]
  & \multirow{2}{*}{$0.255$}
      & $GH_1$ & $\kappa_s = 8.089\times10^{-3},\quad \sigma = 5.300$ \\
  &   & $GH_2$ & $\kappa_s = 9.038\times10^{-3},\quad \sigma = 1.912$ \\[3pt]
  & $0.3$
      & $GH$   & $\kappa_s = 1.126\times10^{-2},\quad \sigma = 2.018$ \\
\hline
\multirow{3}{*}{$(\sigma_i,\,\sigma)$}
  & $0.2$   & $GH$ & $\sigma_i = 1.963,\quad \sigma = 7.661$ \\
  & $0.255$ & $GH$ & $\sigma_i = 1.988,\quad \sigma = 5.003$ \\
  & $0.3$   & $GH$ & $\sigma_i = 2.112,\quad \sigma = 4.843$ \\
\hline\hline
\end{tabular}
\end{table}

In the remainder we treat $(n,d)$ as the primary control knobs: $n$ sets the 
degree of cooperativity and $d$ tunes the net product removal and flux 
balance. Their interplay organizes the Hopf, saddle-node-of-cycles, and isola 
structures described below.

\subsection{Bifurcation curves and codimension-2 organizing centers}

Fig.~\ref{fig2}(a--b) shows the two-parameter bifurcation diagram in the 
$(n,d)$-plane. The Hopf bifurcation curve $H$ decomposes into a supercritical 
branch $H_s$ and a subcritical branch $H_u$, which meet at two generalized 
Hopf (Bautin) points $GH_1$ and $GH_2$. At each generalized Hopf point the 
Hopf criticality changes and a branch of saddle-node bifurcations of limit 
cycles is born.

The generalized Hopf points correspond to parameter values where the first 
Lyapunov coefficient vanishes ($\sigma_1=0$) while the second is nonzero 
($\sigma_2\neq 0$). The sign of $\sigma_2$ determines the stability ordering 
of the two small-amplitude cycles created in the codimension-2 scenario: for 
$\sigma_2<0$ the outer cycle is stable, whereas for $\sigma_2>0$ the outer 
cycle is unstable and an intermediate unstable cycle acts as a separatrix 
between coexisting attractors.

In addition to $H$, we compute a saddle-node-of-limit-cycles curve $SNL$, 
along which a stable and an unstable periodic orbit collide. The $SNL$ curve 
terminates at a codimension-2 cusp-of-cycles point $CPL$, where two 
saddle-node-of-cycles points coalesce. This cusp acts as the global organizing 
center for the creation and annihilation of detached branches of periodic 
solutions (isolas), as described below. The key bifurcation objects are: 
$H_s$/$H_u$ (supercritical/subcritical Hopf bifurcation curves); $GH$ 
(generalized Hopf (Bautin) points, where $\sigma_1=0$ and an $SNL$ curve is 
born); $SNL$ (saddle-node bifurcation curve of limit cycles); and $CPL$ 
(cusp-of-cycles point, codimension~2, on the $SNL$ curve).

\subsection{Regime map in the $(n,d)$-plane}

The global bifurcation diagram in Fig.~\ref{fig2}(a) partitions the 
$(n,d)$-plane into six qualitative regimes (Regions~I--VI), summarized in 
Table~\ref{tab:regimes} and illustrated by representative phase portraits in 
Fig.~\ref{fig4}. The regions differ in the number and stability of coexisting 
attractors and in whether oscillations occur on a connected branch or on an 
isola.

A key feature is the triangular neighborhood near $GH_2$ and $CP_L$, which 
contains the parameter region where disconnected branches of periodic 
solutions appear. Here the system supports three limit cycles arising from 
a single Hopf bifurcation point: two stable orbits of distinct amplitude 
separated by an unstable orbit. This differs from the mechanism reported 
in Ref.~\cite{wang2026}, where three coexisting limit cycles originate from separate 
Hopf points. The $SNL$ branch born at $GH_2$ provides an explicit boundary 
between the two-limit-cycle region and the three-limit-cycle region near 
this codimension-2 cusp of limit cycles -- to our knowledge, the first such 
explicit identification, and a mechanism distinct from that of Ref.~\cite{wang2026}. 
Region I in Fig.~\ref{fig2}(b) is the isola region, in which a family of isolated 
limit-cycle branches originates from the isola center.

\subsection{Representative isolas and their periods}

\begin{figure}[htbp]
    \centering
\includegraphics[width=0.9\columnwidth]{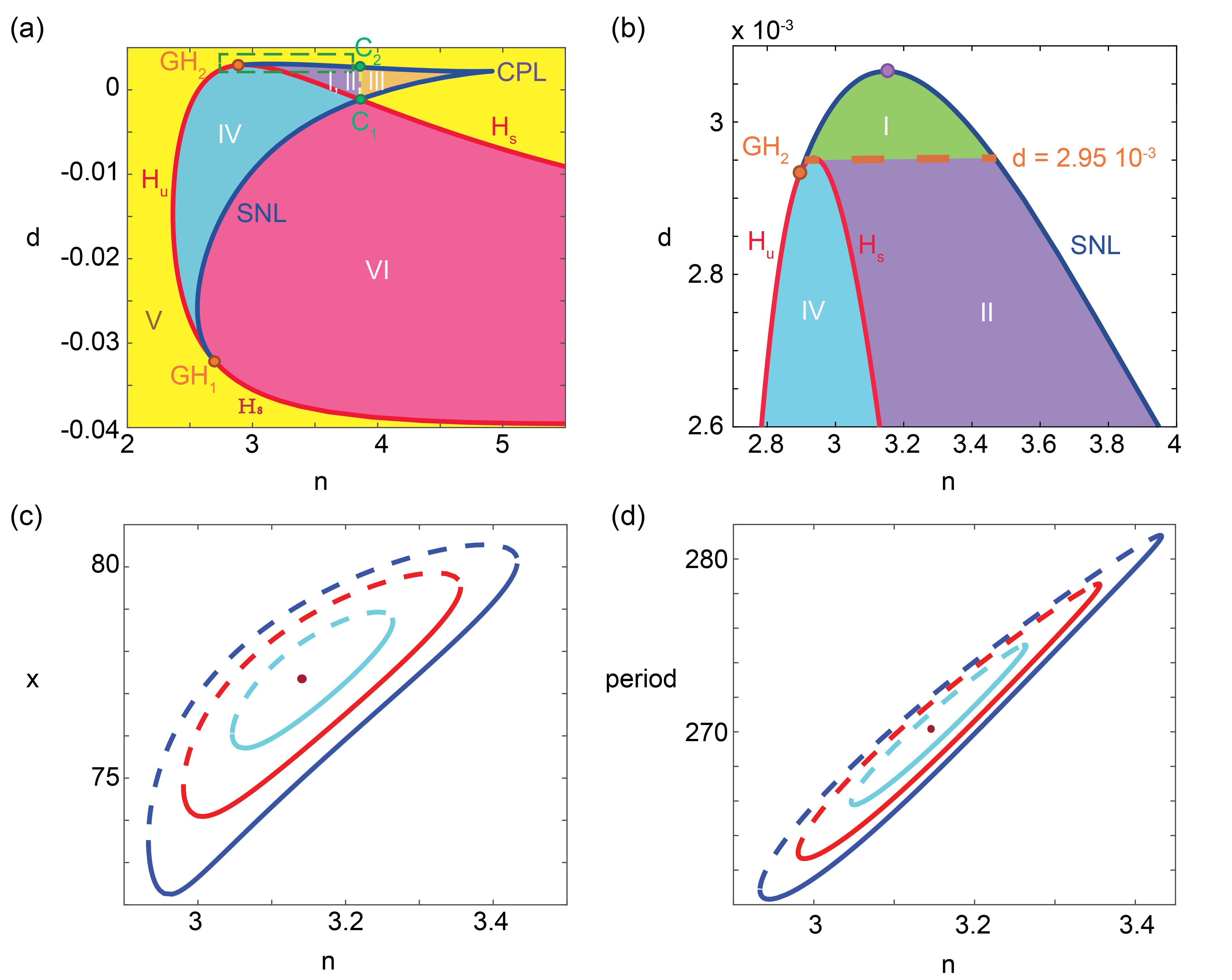}
\caption{(a)~Two-parameter bifurcation diagram of 
system~\eqref{eq:model} in the $(n,d)$-plane. The Hopf curve $H$ decomposes 
into supercritical ($H_s$) and subcritical ($H_u$) branches meeting at 
generalized Hopf points $GH_1$ and $GH_2$. The saddle-node-of-limit-cycles 
curve $SNL$ terminates at the cusp-of-cycles point $CPL$. Together these 
curves partition the plane into six dynamical regimes (Regions~I--VI; see 
Table~\ref{tab:regimes}). $C_1$ denotes the intersection of $H_s$ and $SNL$; $C_2$ denotes the foot 
of the perpendicular dropped from $C_1$ onto the line $d = 2.95\times10^{-3}$.
(b)~Magnified view of the boxed region in panel~(a), showing the isola region 
(Region~I) and the isola center (filled circle).
(c)~Three representative isolas obtained by one-parameter continuation in $n$ 
at fixed values $d = 2.974\times10^{-3}$, $3.014\times10^{-3}$, and 
$3.492\times10^{-3}$. The filled circle marks the isola center.
(d)~Oscillation periods along the three isola branches in panel~(c).}
    \label{fig2}
\end{figure}

\begin{figure}[htbp]
    \centering
\includegraphics[width=0.9\columnwidth]{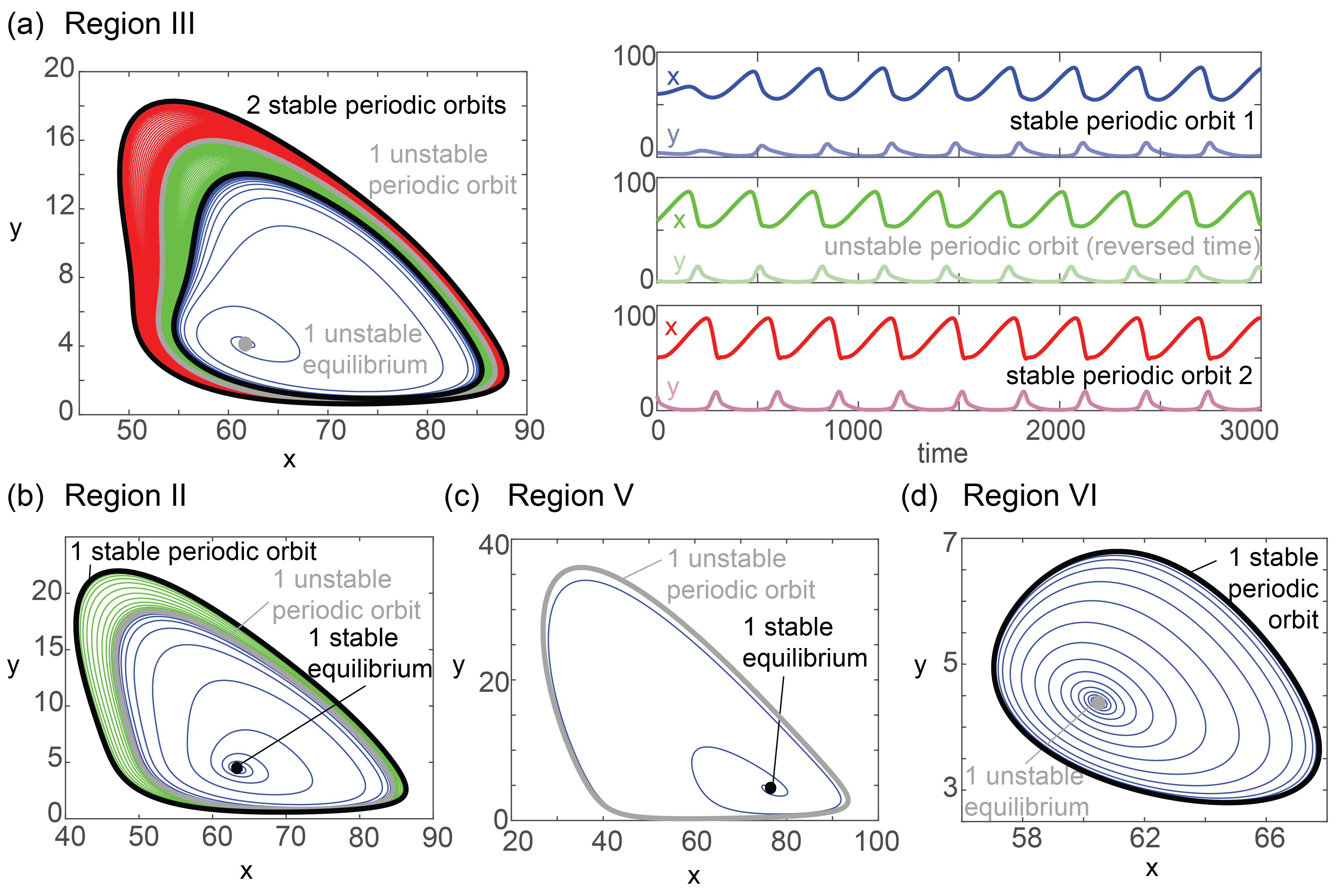}
\caption{Phase portraits illustrating the dynamical regimes of
Table~\ref{tab:regimes}.
(a)~Region~III (birhythmicity): two stable limit cycles coexist with one 
unstable periodic orbit (separatrix) and one unstable equilibrium. Time 
series (right) show $x$ (dark) and $y$ (light) for trajectories converging 
to stable orbit~1 (blue), the unstable orbit in reversed time (green), and 
stable orbit~2 (red).
(b)~Region~II: one stable periodic orbit, one unstable periodic orbit 
(separatrix), and one stable equilibrium.
(c)~Region~V: one unstable periodic orbit and one stable equilibrium; no 
stable oscillation.
(d)~Region~VI: single stable limit cycle born at a supercritical Hopf 
bifurcation, with one unstable equilibrium.}
    \label{fig4}
\end{figure}

Within Region~I, periodic solutions exist on detached branches bounded by two 
saddle-node-of-cycle points on the $SNL$ curve. Fig.~\ref{fig2}(c) shows 
three representative isolas obtained by one-parameter continuation in $n$ at 
fixed values of $d$; the corresponding oscillation periods are shown in 
Fig.~\ref{fig2}(d). As $d$ is varied each isola shrinks and eventually 
collapses at an isola center, marked in Fig.~\ref{fig2}(b--d).

These results show that tuning cooperativity can create oscillations on 
disconnected branches — oscillations with no continuous connection to the 
equilibrium through a Hopf bifurcation. In the context of glycolysis this 
provides a natural mechanism for the abrupt appearance and disappearance of 
rhythmic modes over narrow parameter windows, and sets the stage for 
threshold-dependent switching between oscillatory attractors 
(Section~\ref{sec:time}).

\begin{table}[htbp]
\caption{Summary of dynamical regimes in the $(n,d)$-plane
(Fig.~\ref{fig2}).}
\label{tab:regimes}
\centering
\renewcommand{\arraystretch}{1.1}
\begin{tabular}{l l l}
\hline\hline
Region & Attractors (qualitative) & Outcome \\
\hline
I &
\begin{minipage}[t]{0.58\columnwidth}\raggedright
\textbf{Isola window.} Within the isola parameter interval, a stable
equilibrium, an unstable periodic orbit (outer), and a stable periodic
orbit (inner) on the disconnected branch coexist; outside this interval
only the stable equilibrium remains.
\end{minipage}
&
\begin{minipage}[t]{0.28\columnwidth}\raggedright
Bistable (equilibrium and isolated oscillation)
\end{minipage}
\\[4pt]
II &
\begin{minipage}[t]{0.58\columnwidth}\raggedright
One unstable equilibrium and one stable periodic orbit, with an additional
unstable cycle acting as separatrix.
\end{minipage}
&
\begin{minipage}[t]{0.28\columnwidth}\raggedright
Monostable (oscillation only)
\end{minipage}
\\[4pt]
III &
\begin{minipage}[t]{0.58\columnwidth}\raggedright
\textbf{Birhythmicity.} Two stable periodic orbits (small- and
large-amplitude) coexist, separated by an unstable periodic orbit.
\end{minipage}
&
\begin{minipage}[t]{0.28\columnwidth}\raggedright
Bistable (two oscillations)
\end{minipage}
\\[4pt]
IV &
\begin{minipage}[t]{0.58\columnwidth}\raggedright
Stable equilibrium coexists with a stable periodic orbit (outer), separated
by an unstable periodic orbit (inner).
\end{minipage}
&
\begin{minipage}[t]{0.28\columnwidth}\raggedright
Bistable (equilibrium and oscillation)
\end{minipage}
\\[4pt]
V &
\begin{minipage}[t]{0.58\columnwidth}\raggedright
Stable equilibrium only; no stable periodic orbit.
\end{minipage}
&
\begin{minipage}[t]{0.28\columnwidth}\raggedright
Monostable (equilibrium only)
\end{minipage}
\\[4pt]
VI &
\begin{minipage}[t]{0.58\columnwidth}\raggedright
Unstable equilibrium and a stable periodic orbit born at a supercritical
Hopf bifurcation.
\end{minipage}
&
\begin{minipage}[t]{0.28\columnwidth}\raggedright
Monostable (oscillation only)
\end{minipage}
\\
\hline\hline
\end{tabular}
\end{table}

\section{Time-domain dynamics and switching between coexisting attractors}
\label{sec:time}

We now complement the bifurcation analysis of Section~\ref{sec:global} with 
direct time-domain simulations, illustrating how the dynamical regimes of 
Table~\ref{tab:regimes} manifest in practice. All simulations use the 
baseline parameters of Table~\ref{tab:params} unless otherwise stated.

\subsection{Coexisting oscillatory states in the birhythmicity region}

In Region~III (birhythmicity), two stable periodic attractors of distinct 
amplitude and period coexist, separated by an unstable intermediate limit 
cycle that forms the boundary between their basins of attraction. The 
attractor reached therefore depends on the initial condition.
To demonstrate this, we consider two initial conditions located on 
opposite sides of the unstable periodic orbit. Fig.~\ref{fig4}(a) shows that 
trajectories starting near the equilibrium converge to the small-amplitude 
oscillation, whereas trajectories starting further away converge to the 
large-amplitude oscillation. The time series display clear differences in 
amplitude and period between the two attractors, consistent with their lying 
on separate branches of periodic solutions in the bifurcation diagram. In 
biochemical systems, selection between these states could be influenced by 
transient changes in substrate supply, enzyme activity, or regulatory 
feedback strength.

\subsection{Isolas and threshold-dependent onset of oscillations}

Oscillations in Region~I exist only on a detached branch with no Hopf 
connection to the equilibrium. The dynamical consequence is a 
threshold-like activation mechanism: small perturbations from the 
equilibrium decay back to rest, while sufficiently large perturbations 
reach the basin of attraction of the isolated limit cycle and produce 
sustained oscillations.

This separatrix structure is qualitatively the same as that organizing 
Region~II, shown in Fig.~\ref{fig4}(b): a stable equilibrium and a stable 
periodic orbit are separated by an unstable periodic orbit acting as a 
basin boundary, so that the outcome of a perturbation depends on whether 
it lands inside or outside this unstable orbit. The distinguishing feature 
of Region~I is topological rather than local: the stable equilibrium and 
the stable cycle in Region~II both terminate on the same connected branch 
through a Hopf bifurcation, whereas in Region~I the stable cycle lies on a 
fully disconnected branch bounded by saddle-node-of-cycles points (Fig.~\ref{fig2}b--c), 
with no continuous path to the equilibrium through any Hopf point. 
Consequently, while the onset of oscillation in Region~II can in principle 
be approached continuously by tracking the bifurcation parameter through 
the Hopf point, no such continuous route exists in Region~I: the isolated 
cycle can only be reached by a perturbation large enough to cross the 
unstable outer orbit, making the onset of rhythmic activity abrupt and 
all-or-nothing.

This is in clear contrast to Region~VI (Fig.~\ref{fig4}d), where the single 
stable limit cycle is born directly at a supercritical Hopf bifurcation: 
here the oscillation amplitude grows continuously from zero as the 
bifurcation parameter is varied, with no threshold and no separatrix 
involved. Region~V (Fig.~\ref{fig4}c) provides the complementary 
monostable baseline at the other extreme: only the stable equilibrium 
persists, and the unstable periodic orbit bounds no stable cycle at all, 
so every perturbation decays back to rest regardless of amplitude. Taken 
together, Fig.~\ref{fig4}(c)--(d) bracket the bistable regimes of panels 
(a)--(b): smooth Hopf onset of oscillation on one side (VI), no sustainable 
oscillation on the other (V), with the isola and birhythmic regimes 
occupying the qualitatively distinct bistable middle ground.

Taken together, the control parameters $n$ (cooperativity) and $d$ (net 
product removal) act as biologically meaningful tuning knobs that move the 
system between a quiescent steady state, monorhythmic oscillations, 
birhythmicity, and the isola regime. Cooperativity-induced isolas in 
particular provide a mechanism for switching between metabolic rhythms 
without passing through a classical Hopf transition.

\section{Conclusion}
\label{sec:conclusion}

We analyzed a glycolysis model with allosteric feedback and product recycling
to determine how cooperativity, via the Hill exponent, shapes the global
organization of oscillations. We derived a sharp condition for the existence
of a unique positive equilibrium and characterized its local stability,
establishing Hopf bifurcation criteria for $n=2$ and identifying generalized
Hopf points that organize transitions between oscillatory regimes.

The two-parameter bifurcation diagram in the $(n,d)$-plane provides an
explicit map of six dynamical regimes, organized by a codimension-2
cusp-of-cycles point that governs the creation and annihilation of isolas
and their relationship to birhythmic dynamics.

Numerical continuation shows that Hopf curves can close and that saddle-node
bifurcations of limit cycles together with cusp points generate isolas of
periodic solutions. These isolas produce narrow parameter windows where
oscillations exist on disconnected branches, providing a concrete mechanism
for transitions between mono- and birhythmic dynamics. Tuning cooperativity
or net product removal can create, isolate, or destroy oscillatory states,
yielding coexistence of distinct rhythms consistent with birhythmicity
reported in glycolytic and related biochemical
systems~\cite{biswas2017,pisarchik2014,biswas2023}.

The explicit boundary between the two- and three-limit-cycle regions 
identified near $GH_2$ and $CP_L$ (Section~III\,C) illustrates how 
difficult it generally is to resolve the local structure near a 
codimension-2 cusp of limit cycles; here, continuation directly exposes 
this boundary as an $SNL$ branch.

The dynamical regimes identified here correspond to experimentally accessible
observables in glycolysis, such as oscillations in ATP, ADP, and NADH
concentrations measured via fluorescence or metabolite assays. The coexistence
of multiple limit cycles would manifest as distinct oscillatory modes ---
for example low- versus high-amplitude NADH or ATP oscillations --- that
could be selected by initial conditions or environmental changes.

Our results show that cooperativity is not only a quantitative modifier of
oscillation shape and period, but a key control parameter that reorganizes
the global bifurcation structure of biochemical oscillators. More broadly,
the isola mechanism identified here suggests that cooperative feedback may
be a widespread organizing principle for multistable rhythmicity in metabolic
and regulatory networks --- one that is amenable to detection in experimental time-series data via the kind of 
data-driven approaches recently developed for glycolytic 
oscillators ~\cite{Prokop2024,Prokop2025,CampsValls2023}. Diffusion is also known to play a significant role 
in the bifurcation dynamics of glycolysis models \cite{Wei2019,Zhou2015,MojicaBenavides2021,Bashkirtseva2023}; extending the 
present cusp-of-cycles and isola analysis to a spatially extended, 
reaction--diffusion setting is a natural direction for future work.

\medskip
\noindent\textbf{Code availability.}\;
AUTO-07P continuation codes for the bifurcation diagrams are available at
\url{https://github.com/YoungTarzan/Glycolysis_codes}.

\section*{Acknowledgment}
The authors are grateful to Professor Sandip Kar and John J. Tyson for their helpful suggestions. This project is supported by NSF of Zhejiang (LY20A010002).


%
%

%


\clearpage
\appendix
\section{Existence and uniqueness of the positive equilibrium}
\label{app:equilibrium}

Setting the right-hand sides of \eqref{eq:model} equal to zero yields the 
algebraic system
\begin{equation}
\begin{aligned}
0 &= \nu - \frac{\sigma\,x(x+1)(y+1)^2}{L + (x+1)^2 (y+1)^2} 
+ \frac{\sigma_i y^n}{\kappa^n + y^n},\\
0 &= \frac{q\sigma\,x(x+1)(y+1)^2}{L + (x+1)^2 (y+1)^2} 
-\frac{q\sigma_i y^n}{\kappa^n + y^n} - (\kappa_s+d) y.
\end{aligned}
\label{eq:equil_system}
\end{equation}
Multiplying the first equation in \eqref{eq:equil_system} by $q$ and adding 
it to the second eliminates the nonlinear flux terms, giving
\[
(\kappa_s+d)\,y^* = q\nu \qquad\Longrightarrow\qquad 
y^*=\frac{q\nu}{\kappa_s+d},
\]
so $y^*>0$ whenever $\kappa_s+d>0$.

Substituting $y^*$ into \eqref{eq:equil_system} and solving the resulting 
quadratic equation yields an explicit expression for the equilibrium,
\begin{equation}
x^*=\frac{\kappa^n M+(y^*)^n N}
{2(1+y^*)^2\left(\kappa^n(\nu-\sigma)+(y^*)^n(\nu-\sigma+\sigma_i)\right)},
\qquad
y^*=\frac{q\nu}{\kappa_s+d},
\label{eq:equil_explicit}
\end{equation}
which is valid under condition~\eqref{eq:equilibrium_condition}, i.e.\ when 
$A_2 < 0$. Here
\begin{align*}
M&=\sigma-2\nu(y^*)^2-4\nu y^*-2\nu+\sigma(y^*)^2+2\sigma y^*+\rho,\\
N&=\sigma-2\sigma_i-2\nu(y^*)^2-4\nu y^*+2\nu+\sigma(y^*)^2
-2\sigma_i (y^*)^2+2\sigma y^*-4\sigma_i y^*+\rho,\\
\rho&=\sqrt{\sigma^2(1+y^*)^4-\frac{4L(1+y^*)^2
\big(\nu\kappa^n+(y^*)^n(\sigma_i+\nu)\big)
\big(\kappa^n(\nu-\sigma)+(y^*)^n(-\sigma+\sigma_i+\nu)\big)}
{\big(\kappa^n+(y^*)^n\big)^2}}.
\end{align*}

An equivalent algebraic route is to eliminate $y$ first. Substituting 
$y^*=\frac{q\nu}{\kappa_s+d}$ into the first equation of 
\eqref{eq:equil_system} yields a quadratic equation in $x^*$,
\begin{equation}
F(x^*)=A_2(x^*)^2+A_1x^*+A_0=0,
\label{eq:Fx}
\end{equation}
with coefficients
\begin{align*}
A_2&=\frac{(\kappa_s+d+q\nu)^2\Big((\nu-\sigma)\kappa^n(\kappa_s+d)^n
+(\nu-\sigma+\sigma_i)q^n\nu^n\Big)}
{(\kappa_s+d)^2\Big(\kappa^n(\kappa_s+d)^n+q^n\nu^n\Big)},\\[4pt]
A_1&=\frac{(\kappa_s+d+q\nu)^2\Big((2\nu-\sigma)\kappa^n(\kappa_s+d)^n
+(2\nu-\sigma+2\sigma_i)q^n\nu^n\Big)}
{(\kappa_s+d)^2\Big(\kappa^n(\kappa_s+d)^n+q^n\nu^n\Big)},\\[4pt]
A_0&=\frac{\big(d^2L+d^2+2d\kappa_s L+2d\kappa_s+2dq\nu+\kappa_s^2 L
+\kappa_s^2+2\kappa_s q\nu+q^2\nu^2\big)\,\zeta}
{(\kappa_s+d)^2\Big(\kappa^n(\kappa_s+d)^n+q^n\nu^n\Big)},
\end{align*}
where
\[
\zeta=\nu\kappa^n(\kappa_s+d)^n+\sigma_i q^n\nu^n+\nu q^n\nu^n.
\]
In particular, $A_0>0$. By Descartes' rule of signs, \eqref{eq:Fx} admits 
at most one positive root (Table~\ref{tab:equilibrium_signs}), and a positive 
equilibrium exists if and only if $A_2<0$, i.e.,
\begin{equation}
\sigma>\nu+\frac{\sigma_i q^n\nu^n}{\kappa^n(\kappa_s+d)^n+q^n\nu^n}.
\label{eq:equilibrium_condition}
\end{equation}
This establishes existence and uniqueness of the positive equilibrium in the 
biologically relevant region.

\begin{table}[htbp]
\centering
\caption{Existence of a positive equilibrium based on the signs of 
$(A_2,A_1,A_0)$ in \eqref{eq:Fx}.}
\label{tab:equilibrium_signs}
\begin{tabular}{cccc}
\toprule
Sign of $A_2$ & Sign of $A_1$ & Sign of $A_0$ & 
Number of positive equilibria\\
\midrule
$>0$ & $>0$ & $>0$ & None\\
$>0$ & $<0$ & $>0$ & None\\
$<0$ & $>0$ & $>0$ & One\\
$<0$ & $<0$ & $>0$ & One\\
\bottomrule
\end{tabular}
\end{table}

\section{Linearization and Hopf bifurcation condition}
\label{app:hopf}

Let $E^*=(x^*,y^*)$ be the (unique) positive equilibrium. The Jacobian 
matrix of \eqref{eq:model} at $E^*$ is
\[
J(E^*)=\begin{pmatrix}
a_{10} & a_{01}\\
b_{10} & b_{01}
\end{pmatrix},
\]
where
\begin{align*}
a_{10}&=-\frac{\sigma(1+y^*)^2\Big((1+2x^*)L+(1+x^*)^2(1+y^*)^2\Big)}
{\Big(L+(1+x^*)^2(1+y^*)^2\Big)^2},\\
a_{01}&=-\frac{2\sigma x^*(1+x^*)(1+y^*)L}
{\Big(L+(1+x^*)^2(1+y^*)^2\Big)^2}
+\frac{n\sigma_i\kappa^n (y^*)^{n-1}}{\Big(\kappa^n+(y^*)^n\Big)^2},\\
b_{10}&=\frac{q\sigma(1+y^*)^2\Big((1+2x^*)L+(1+x^*)^2(1+y^*)^2\Big)}
{\Big(L+(1+x^*)^2(1+y^*)^2\Big)^2},\\
b_{01}&=\frac{2q\sigma x^*(1+x^*)(1+y^*)L}
{\Big(L+(1+x^*)^2(1+y^*)^2\Big)^2}
-\frac{nq\sigma_i\kappa^n (y^*)^{n-1}}
{\Big(\kappa^n+(y^*)^n\Big)^2}-(\kappa_s+d).
\end{align*}

A direct computation yields
\[
\det(J(E^*))=a_{10}b_{01}-a_{01}b_{10}
=\frac{\sigma(1+y^*)^2\Big((1+2x^*)L+(1+x^*)^2(1+y^*)^2\Big)}
{\Big(L+(1+x^*)^2(1+y^*)^2\Big)^2}(\kappa_s+d)>0
\]
whenever $\kappa_s+d>0$. Therefore, local stability of $E^*$ is controlled 
by the trace $\operatorname{tr}(J(E^*))=a_{10}+b_{01}$. In addition, 
$\det(J(E^*))>0$ rules out a Bogdanov--Takens bifurcation.

A Hopf bifurcation occurs when
\[
\operatorname{tr}(J(E^*))=0,\qquad \det(J(E^*))>0.
\]
The sign of the first Lyapunov coefficient $\sigma_1$ determines whether 
the Hopf bifurcation is supercritical ($\sigma_1<0$) or subcritical 
($\sigma_1>0$). At generalized Hopf points ($\sigma_1=0$ and 
$\sigma_2\neq 0$), the sign of $\sigma_2$ determines the local two-cycle 
configuration (Section~\ref{sec:global} and Appendix~\ref{app:lyapunov}).

\section{Lyapunov coefficients and degenerate Hopf bifurcation}
\label{app:lyapunov}

The first Lyapunov coefficient $\sigma_1$ and second Lyapunov coefficient 
$\sigma_2$ are computed using the algorithm of Yu~\cite{Yu1998}. The Taylor 
expansion coefficients and transformed normal form coefficients are listed 
below.

\subsection{Taylor expansion coefficients}
$$
\begin{aligned}
a_{10} &= -\frac{\sigma}{\left(L + (x_*+1)^2 (y_*+1)^2\right)^2} (y_*+1)^2 \left(2 L x_* + L + (x_*+1)^2 (y_*+1)^2\right), \\
a_{11} &= -\frac{\sigma}{\left(L + (x_*+1)^2 (y_*+1)^2\right)^3} 2 L (y_*+1) \left(2 L x_* + L - (x_*+1)^2 (2 x_* - 1) (y_*+1)^2\right), \\
a_{12} &= -\frac{\sigma}{\left(L + (x_*+1)^2 (y_*+1)^2\right)^4} L \left(L^2 (2 x_* + 1) - 2 L (x_*+1)^2 (8 x_* + 1) (y_*+1)^2
\right. \\
&\quad \left. + 3 (x_*+1)^4 (2 x_* - 1) (y_*+1)^4\right), \\
a_{13} &= -\frac{\sigma}{\left(L + (x_*+1)^2 (y_*+1)^2\right)^5} 4 L (x_*+1)^2 (y_*+1) \left(L^2 (4 x_* + 1) - 10 L x_* (x_*+1)^2 (y_*+1)^2 \right. \\
&\quad \left. + (x_*+1)^4 (2 x_* - 1) (y_*+1)^4\right), \\
a_{20} &= \frac{\sigma}{\left(L + (x_*+1)^2 (y_*+1)^2\right)^3} \left(-L^2 (y_*+1)^2 + 3 L x_* (x_*+1) (y_*+1)^4 + (x_*+1)^3 (y_*+1)^6\right),\\
a_{21} &= \frac{\sigma}{\left(L + (x_*+1)^2 (y_*+1)^2\right)^4} 2 L (y_*+1) \left(L^2 - 2 L \left(4 (x_*)^2 + 5 x_* + 1\right) (y_*+1)^2
 \right. \\
&\quad \left. +
3 (x_*-1) (x_*+1)^3 (y_*+1)^4\right), \\
a_{22} &= \frac{\sigma}{\left(L + (x_*+1)^2 (y_*+1)^2\right)^5} L \left(L^3 - L^2 (13 + 44 x_* + 31 (x_*)^2) (1 + y_*)^2 \right. \\
&\quad \left. + 5 L (1 + x_*)^3 (-1 + 11 x_*) (1 + y_*)^4 - 9 (-1 + x_*) (1 + x_*)^5 (1 + y_*)^6\right), \\
a_{30} &= \frac{\sigma}{\left(L + (x_*+1)^2 (y_*+1)^2\right)^4} \left(-L^2 (4 x_* + 3) (y_*+1)^4 + 2 L (x_*+1)^2 (2 x_* - 1) (y_*+1)^6 \right. \\
&\quad \left. + (x_*+1)^4 (y_*+1)^8\right),\\
a_{31} &= \frac{\sigma}{\left(L + (x_*+1)^2 (y_*+1)^2\right)^5} 4 L (y_*+1)^3 \left(-L^2 (4 x_* + 3) + 10 L x_* (x_*+1)^2 (y_*+1)^2 \right. \\
&\quad \left. - (x_*+1)^4 (2 x_* - 3) (y_*+1)^4\right), \\
a_{40} &= \frac{\sigma}{\left(L + (x_*+1)^2 (y_*+1)^2\right)^5} (y_*+1)^4 \left(L^3 - 5 L^2 (x_*+1) (2 x_* + 1) (y_*+1)^2 \right. \\
&\quad \left. + 5 L (x_*-1) (x_*+1)^3 (y_*+1)^4 + (x_*+1)^5 (y_*+1)^6\right), \\
a_{41} &= \frac{\sigma}{\left(L + (x_*+1)^2 (y_*+1)^2\right)^6} 2 L (y_*+1)^3 \left(-2 L^3 + 3 L^2 \left(11 (x_*)^2 + 17 x_* + 6\right) (y_*+1)^2 \right. \\
&\quad \left. - 10 L (x_*+1)^3 (4 x_* - 1) (y_*+1)^4 + 5 (x_*-2) (x_*+1)^5 (y_*+1)^6\right), \\
a_{01} &= \frac{2 \kappa^2 y_* \sigma_i}{\left(\kappa^2 + (y_*)^2\right)^2} - \frac{2 L \sigma x_* (x_*+1) (y_*+1)}{\left(L + (x_*+1)^2 (y_*+1)^2\right)^2}, \\
a_{02} &= \frac{\kappa^2 \sigma_i \left(\kappa^2 - 3 (y_*)^2\right)}{\left(\kappa^2 + (y_*)^2\right)^3} + \frac{L \sigma x_* (x_*+1) \left(3 (x_*+1)^2 (y_*+1)^2 - L\right)}{\left(L + (x_*+1)^2 (y_*+1)^2\right)^3}, \\\end{aligned}
$$

\begin{align*}
a_{03} &= \frac{4 \kappa^2 y_* \sigma_i \left((y_*)^2 - \kappa^2\right)}{\left(\kappa^2 + (y_*)^2\right)^4} - \frac{4 L \sigma x_* (x_*+1)^3 (y_*+1) \left((x_*+1)^2 (y_*+1)^2 - L\right)}{\left(L + (x_*+1)^2 (y_*+1)^2\right)^4}, \\
a_{04} &= \frac{L \sigma x_* (x_*+1)^3 \left(L^2 - 10 L (x_*+1)^2 (y_*+1)^2 + 5 (x_*+1)^4 (y_*+1)^4\right)}{\left(L + (x_*+1)^2 (y_*+1)^2\right)^5} \\
&\quad - \frac{\kappa^2 \sigma_i \left(\kappa^4 + 5 (y_*)^4 - 10 \kappa^2 (y_*)^2\right)}{\left(\kappa^2 + (y_*)^2\right)^5},
\end{align*}
$b_{10}=q a_{10}$, $b_{11}=q a_{11}$, $b_{12}=q a_{12}$, $b_{13}=q a_{13}$, $b_{20}=q a_{20}$, $b_{21}=q a_{21}$, $b_{22}=q a_{22}$, $b_{30}=q a_{30}$,\\
$b_{31}=q a_{31}$, $b_{40}=q a_{40}$, $b_{01}=q a_{01}-d-\kappa_s$, $b_{02}=q a_{02}$, $b_{03}=q a_{03}$, $b_{04}=q a_{04}$.
\begin{align*}
A_{11} &= \frac{-\left(2 a_{02} a_{10} - a_{01} a_{11}\right)}{a_{01}}, \quad
A_{12} = -\frac{\left(a_{01} a_{12} \lambda - 3 a_{03} a_{10} \lambda \right)}{a_{01}}, \\
A_{13} &= -\frac{\left(4 a_{04} a_{10} \lambda^2 - a_{01} a_{13} \lambda^2\right)}{a_{01}}, \quad
A_{20} = \frac{-a_{20} a_{01}^2 + a_{10} a_{11} a_{01} - a_{02} a_{10}^2}{a_{01} \lambda}, \\
A_{21} &= \frac{a_{21} a_{01}^2 - 2 a_{10} a_{12} a_{01} + 3 a_{03} a_{10}^2}{a_{01}}, \quad
A_{22}= \frac{-6 a_{04} a_{10}^2 \lambda + 3 a_{01} a_{10} a_{13} \lambda - a_{01}^2 a_{22} \lambda}{a_{01}}, \\
A_{30} &= \frac{a_{30} a_{01}^3 - a_{10} a_{21} a_{01}^2 + a_{10}^2 a_{12} a_{01} - a_{03} a_{10}^3}{a_{01} \lambda},\\
A_{31} &= \frac{-a_{31} a_{01}^3 + 2 a_{10} a_{22} a_{01}^2 - 3 a_{10}^2 a_{13} a_{01} + 4 a_{04} a_{10}^3}{a_{01}}, \\
A_{40} &= \frac{a_{31} a_{01}^3 (-\lambda) + 2 a_{10} a_{22} a_{01}^2 \lambda - 3 a_{10}^2 a_{13} a_{01} \lambda + 4 a_{04} a_{10}^3 \lambda}{a_{01} \lambda},\\
A_{02} &=-\frac{a_{02} \lambda }{a_{01}}, \quad A_{03} =\frac{a_{03} \lambda ^2}{a_{01}}, \quad A_{04} =-\frac{a_{04} \lambda ^3}{a_{01}}\\
B_{11} &= \frac{a_{01} b_{11} - 2 a_{10} b_{02}}{\lambda}, \quad B_{12} = 3 a_{10} b_{03} - a_{01} b_{12}, \\
B_{13} &= a_{01} b_{13} \lambda - 4 a_{10} b_{04} \lambda, \quad B_{20} = \frac{a_{01}^2 b_{20} - a_{10} a_{01} b_{11} + a_{10}^2 b_{02}}{\lambda^2}, \\
B_{21} &= \frac{a_{01}^2 \left(-b_{21}\right) + 2 a_{10} a_{01} b_{12} - 3 a_{10}^2 b_{03}}{\lambda}, \quad B_{22} = a_{01}^2 b_{22} - 3 a_{10} a_{01} b_{13} + 6 a_{10}^2 b_{04}, \\
B_{30} &= \frac{a_{01}^3 \left(-b_{30}\right) + a_{10} a_{01}^2 b_{21} - a_{10}^2 a_{01} b_{12} + a_{10}^3 b_{03}}{\lambda^2}, \\
B_{31} &= \frac{a_{01}^3 b_{31} - 2 a_{10} a_{01}^2 b_{22} + 3 a_{10}^2 a_{01} b_{13} - 4 a_{10}^3 b_{04}}{\lambda}, \\
B_{40} &= \frac{a_{01}^4 b_{40} - a_{10} a_{01}^3 b_{31} + a_{10}^2 a_{01}^2 b_{22} - a_{10}^3 a_{01} b_{13} + a_{10}^4 b_{04}}{\lambda^2},\\
B_{01} &= -\frac{b_{01}}{\lambda}, \quad B_{02} = b_{02}, \quad B_{03} = -b_{03}\lambda, \quad B_{04} = b_{04} \lambda^2.
\end{align*}
\small

\subsection*{Second Lyapunov coefficient}
\begin{align*}
\sigma_2 &= \frac{1}{6912}\Bigg[
\left(-480 A_{11}+864 B_{02}+960 B_{20}\right) A_{20}^3
+\left(480 \left(-2 A_{11}+B_{02}+\frac{B_{20}}{5}\right) A_{02}+B_{11} \left(24 A_{11}\right.\right.\\
&\quad \left.\left.-240 B_{02}-432 B_{20}\right)\right)A_{20}^2 +\left(912 A_{12}+720 B_{03}-384 B_{21}+1440\right) A_{20}^2
+\Big(-120 A_{11}^3+\left(312 B_{02}\right.\\
&\quad\left.+216 B_{20}\right)\quad A_{11}^2 +\left(-960 A_{02}^2+240 B_{02}^2+432 B_{20}^2+192 B_{02} B_{20}\right) A_{11}
-864 B_{02}^3+960 B_{20}^3+960 B_{02} B_{20}^2\\
&\quad-960 A_{02}^2 B_{02}\quad +\left(-480 A_{02}^2-480 B_{02}^2\right) B_{20}-192 A_{02} B_{11} \left(\frac{9 A_{11}}{4}+B_{02}+B_{20}\right)
+B_{11}^2 \left(24 A_{11}-312 B_{02}\right.\\
&\quad\left.-216 B_{20}\right)\Big) A_{20}\quad +\Big(2160 \left(\frac{5 A_{12}}{9}+B_{03}-\frac{B_{21}}{9}+\frac{1}{15}\right) A_{02}
+B_{11} \left(408 A_{12}+504 B_{03}-168 B_{21}-72\right)\\
&\quad+A_{11} \left(792 A_{03}\right.\quad \left.+504 A_{21}+360 B_{30}+18\right)
+B_{02} \left(1152 A_{03}-1440 A_{21}+1152 B_{30}-1440\right)+B_{20}\\
&\quad \left(432 A_{03}-1008 A_{21}\right.\quad\left.+1872 B_{30}-612\right)\Big) A_{20}
+\left(-1008 A_{13}-1296 A_{31}-3168 B_{04}-576 B_{22}+864 B_{40}\right) \\
&\quad A_{20}+310 \Bigg(\frac{A_{11}^4}{310}+\left(\frac{11 B_{02}}{155}-\frac{12 B_{20}}{155}\right) A_{11}^3
+\left(-\frac{19}{31} A_{02}^2+\frac{4 B_{02}^2}{31}+\frac{8 B_{20}^2}{155}-\frac{92 B_{02} B_{20}}{155}\right) A_{11}^2 \\
&\quad -\frac{377}{155} B_{20} \left(A_{02}^2+\frac{100 B_{02}^2}{377}+\frac{48 B_{20}^2}{377}+\frac{20 B_{02} B_{20}}{377}\right) A_{11}
+\left(\frac{32 A_{02}}{5}+\frac{1461 B_{11}}{155}\right) A_{20}^3-\frac{17}{155} A_{02} B_{11}^3 \\
&\quad +\Bigg(\frac{1682 A_{02}^2}{155}+\frac{2558 B_{11} A_{02}}{155}+\frac{662 B_{11}^2}{155}+\frac{358 B_{20}^2}{155}
+\left(-\frac{12 B_{02}}{31}-\frac{1257 B_{20}}{155}\right) A_{11}-\frac{8 B_{02}^2}{5}-\frac{528 A_{11}^2}{155} \\
&\quad -\frac{448 B_{02} B_{20}}{155}\Bigg) A_{20}^2
+\Bigg(A_{02}^2+\frac{28 B_{02}^2}{155}+\frac{96 B_{20}^2}{155}
+\left(\frac{119 B_{02}}{155}+\frac{99 B_{20}}{155}\right)
A_{11}+\frac{36 B_{02} B_{20}}{31}-\frac{19 A_{11}^2}{310}\Bigg)\\
&\quad B_{11}^2-\frac{70}{31} B_{20}^2 \left(A_{02}^2+\frac{92 B_{02}^2}{175}-\frac{76 B_{20}^2}{175}-\frac{16 B_{02} B_{20}}{175}\right)
+\Bigg(\frac{3 B_{11}^3}{155}+\frac{107}{31} A_{02} B_{11}^2
+\left(\frac{1669 A_{02}^2}{155}+\frac{216 B_{20}^2}{155}\right.\\
&\quad\left.+\left(\frac{8 B_{02}}{5}-\frac{184 B_{20}}{155}\right) A_{11} \right. \left. +\frac{8 B_{02} B_{20}}{155}-\frac{192 A_{11}^2}{155}-\frac{172 B_{02}^2}{155}\right) B_{11}
-\frac{64}{155} A_{02} \left(\frac{705 A_{11}^2}{64}+\left(\frac{13 B_{02}}{2}\right.\right.\\
&\quad\left.\left. +\frac{637 B_{20}}{32}\right) A_{11} \right.
\left. +\left(B_{02}+\frac{25 B_{20}}{4}\right) B_{20}\right)\Bigg) A_{20}+\frac{32}{155} A_{02} B_{11} \left(-\frac{179}{32} A_{11}^2+\left(-\frac{5 B_{02}}{2}-\frac{13 B_{20}}{2}\right) A_{11} \right. \\
&\quad \left. +\left(B_{02}-\frac{15 B_{20}}{8}\right) B_{20}\right)-\frac{74 A_{20}^4}{31}-\frac{2 B_{11}^4}{31}\Bigg) B_{01}^3
+1440 \left(\frac{A_{12}}{3}+B_{03}-\frac{1}{2}-\frac{B_{21}}{6}\right) A_{02}^2 -960 A_{02} \left(\frac{A_{11}}{2}\right.\\
&\quad\left. +B_{02}\right) \Bigg(A_{02}^2+\frac{A_{11}^2}{4}+B_{02}^2
+\left(-\frac{19 B_{02}}{20}-\frac{B_{20}}{4}\right) A_{11}+\frac{B_{02} B_{20}}{10}-\frac{B_{20}^2}{2}\Bigg)-216 A_{02} B_{11}^2 \left(-\frac{A_{11}}{9}+B_{02}\right.\\
&\quad\left.+\frac{5 B_{20}}{9}\right)-1508 B_{01}^2 \Bigg[\left(\frac{573 A_{11}}{377}-\frac{1545 B_{02}}{377}-\frac{1266 B_{20}}{377}\right) A_{20}^3 +\left(B_{11} \left(-\frac{7 A_{11}}{13}+\frac{289 B_{02}}{377}+\frac{966 B_{20}}{377}\right) \right.\\
&\quad\left. -\frac{1480}{377} A_{02} \left(-\frac{251 A_{11}}{370}+B_{02}-\frac{9 B_{20}}{296}\right)\right) A_{20}^2+\Bigg(\frac{83 A_{11}^3}{377}+\left(-\frac{671 B_{02}}{754}-\frac{202 B_{20}}{377}\right) A_{11}^2+\left(\frac{657 A_{02}^2}{377}\right.
\end{align*}
\begin{align*}
&\quad\left.-\frac{140 B_{02}^2}{377}-\frac{837 B_{20}^2}{377}-\frac{209 B_{02} B_{20}}{377}\right) A_{11}+\frac{172 B_{02}^3}{377}+\left(-\frac{97 A_{11}}{377}+\frac{387 B_{02}}{754}+\frac{214 B_{20}}{377}\right) B_{11}^2\\
&\quad-\frac{708}{377} B_{02} B_{20}^2-\frac{1253}{377} A_{02}^2 B_{02}-\frac{708}{377} B_{02} B_{20}^2-\frac{1253}{377} A_{02}^2 B_{02}
+\frac{635}{377} \left(-\frac{2 A_{11}}{635}+B_{02}+\frac{1501 B_{20}}{635}\right)\\
&\quad A_{02} B_{11}+\left(-\frac{62}{377} A_{02}^2-\frac{136 B_{02}^2}{377}\right) B_{20}-\frac{402 B_{20}^3}{377}\Bigg) A_{20}+\left(\frac{105 A_{11}}{754}+\frac{35 B_{02}}{377}+\frac{47 B_{20}}{377}\right) B_{11}^3+\frac{3}{13}\\
&\quad \left(-\frac{4 A_{11}}{87}\right.\left.+B_{02}-\frac{445 B_{20}}{174}\right) A_{02} B_{11}^2+\Bigg(-\frac{21}{754} A_{11}^3+\left(-\frac{229 B_{02}}{377}-\frac{187 B_{20}}{377}\right) A_{11}^2+\left(-\frac{3}{377} A_{02}^2\right.\\
&\quad\left.-\frac{186 B_{02}^2}{377}\right.\left.-\frac{34 B_{20}^2}{377}\right.\left.-\frac{388 B_{02} B_{20}}{377}\right) A_{11}+\frac{96 B_{20}^3}{377}-\frac{4}{29} B_{02} B_{20}^2-\frac{230}{377} A_{02}^2 B_{02}+\left(\frac{7 A_{02}^2}{377}-\frac{12 B_{02}^2}{29}\right) B_{20}\\
&\quad-\frac{56 B_{02}^3}{377}\Bigg) B_{11}+A_{02}\Bigg(\frac{74 A_{11}^3}{377}+\left(\frac{75 B_{02}}{377}+\frac{309 B_{20}}{754}\right) A_{11}^2+\left(\frac{190 A_{02}^2}{377}+\frac{40 B_{02}^2}{377}-B_{20}^2+\frac{179 B_{02} B_{20}}{377}\right)\\
&\quad A_{11}+\left(A_{02}^2+\frac{68 B_{02}^2}{377}\right.\left.+\frac{250 B_{02} B_{20}}{377}-\frac{972 B_{20}^2}{377}\right) B_{20}\Bigg)\Bigg]-760 B_{01} \Bigg[A_{02}^4-B_{02}^2 A_{02}^2+\frac{7}{76} B_{11}^3 A_{02}+\frac{717}{380}\\
&\quad \Bigg(A_{02}^2+\frac{14 A_{11}^2}{717}+\frac{212 B_{02}^2}{239}+\frac{304 B_{20}^2}{717}+\left(\frac{100 B_{02}}{717}+\frac{82 B_{20}}{239}\right) A_{11}+\frac{860 B_{02} B_{20}}{717}\Bigg) B_{11} A_{02}+\frac{3 A_{11}^4}{380}\\
&\quad+\frac{274 A_{20}^4}{95}+\frac{81 B_{11}^4}{760}+\frac{109 B_{20}^4}{38}+\left(-\frac{109 B_{02}}{190}-\frac{23 B_{20}}{38}\right) A_{11}^3 +\left(\frac{667 A_{02}}{95}-\frac{141 B_{11}}{95}\right) A_{20}^3+\frac{47}{19} B_{02} B_{20}^3\\
&\quad+\left(\frac{541 A_{02}^2}{760}-\frac{87 B_{20}^2}{190}-\frac{369 B_{02}^2}{380}\right.\left.-\frac{591 B_{02}| B_{20}}{380}\right) A_{11}^2+\Bigg(\frac{768 A_{02}^2}{95}+\frac{441 B_{11} A_{02}}{380}+\frac{923 A_{11}^2}{760}+\frac{12 B_{02}^2}{95}\\
&\quad+\frac{45 B_{11}^2}{152}+\frac{1669
B_{20}^2}{190}+\left(-\frac{2173 B_{02}}{380}\right.\left.-\frac{339 B_{20}}{190}\right) A_{11}+\frac{295 B_{02} B_{20}}{38}\Bigg) A_{20}^2+\Bigg(\frac{231 A_{02}^2}{760}+\frac{3 A_{11}^2}{152}+\frac{13 B_{02}^2}{19}\\
&\quad+\frac{469 B_{20}^2}{380}+\left(\frac{157 B_{02}}{380}+\frac{59 B_{20}}{76}\right) A_{11}+\frac{87 B_{02} B_{20}}{76}\Bigg) B_{11}^2+\left(-\frac{237}{190} A_{02}^2-\frac{69 B_{02}^2}{190}\right) B_{20}^2+\Bigg(-\frac{96}{95} B_{02}^3\\
&\quad-\frac{861}{380} A_{02}^2 B_{02}-\frac{3}{190} B_{20}^2 B_{02}+\frac{933 B_{20}^3}{380}+\left(-\frac{47}{190} A_{02}^2-\frac{891 B_{02}^2}{380}\right) B_{20}\Bigg) A_{11}+\Bigg(-\frac{37}{380} B_{11}^3+\frac{42}{95} A_{02} B_{11}^2\\
&\quad+\left(\frac{813 A_{02}^2}{190}+\left(\frac{93 B_{02}}{38}+\frac{311 B_{20}}{95}\right) A_{11}\right.\left.-\frac{13 A_{11}^2}{95} \right.\left. -\frac{377 B_{02} B_{20}}{190}-\frac{71 B_{02}^2}{380}-\frac{771 B_{20}^2}{380}\right) B_{11}+\frac{100}{19} \Bigg(A_{02}^2\\
&\quad+\frac{3 A_{11}^2}{8}+\frac{841 B_{02}^2}{1000}+\frac{221 B_{20}^2}{1000}+\left(\frac{29 B_{20}}{250}\right.\left.-\frac{1123 B_2}{1000}\right) A_{11}+\frac{393 B_{02} B_{20}}{250}\Bigg) A_{02}\Bigg) A_{20}+\Bigg(-\frac{78}{95} \left(B_{21}\right.\\
&\quad\left.+\frac{75}{13}\right) A_{02}^2+\frac{216}{95} B_{02} \left(\frac{A_{21}}{3}+B_{30}\right) A_{02}+\left(-\frac{39 A_{12}}{380}+\frac{21 B_{21}}{760}-\frac{9}{95}-\frac{63 B_{03}}{760}\right) A_{11}^2+\left(-\frac{222 A_{12}}{95}\right.\\
&\quad\left.+\frac{3069 B_{03}}{380}+\frac{549 B_{21}}{380}-\frac{1557}{190}\right) A_{20}^2+\left(-\frac{24 B_{21}}{95}-\frac{72}{95}\right) B_{02}^2+\left(-\frac{21 A_{12}}{95}+\frac{9}{10}-\frac{261 B_{03}}{380}-\frac{93 B_{21}}{380}\right) B_{11}^2\\
&\quad+\left(-\frac{42 A_{12}}{95}+\frac{9}{95}-\frac{99 B_{03}}{190}\right.\left.-\frac{27 B_{21}}{190}\right) B_{20}^2+\Bigg(-\frac{153}{190} \left(A_{03}+\frac{81 B_{30}}{68}+\frac{37}{68}-\frac{5 A_{21}}{102}\right) A_{02}+\left(-\frac{21 A_{12}}{76}\right.\\
&\quad\left.+\frac{18 B_{03}}{95}-\frac{9 B_{21}}{190}-\frac{333}{380}\right)B_{02}+\left(\frac{9 A_{12}}{95}+\frac{54 B_{03}}{95}+\frac{12 B_{21}}{19}-\frac{27}{95}\right) B_{20}\Bigg) A_{11}+\Bigg(\frac{1953}{380} \left(-\frac{58 A_{12}}{651}+B_{03}\right.
\end{align*}
\begin{align*}
&\quad\left.-\frac{488}{217}-\frac{25 B_{21}}{651}\right) A_{02}+\left(-\frac{819 A_{03}}{380}+\frac{3 A_{21}}{19}-\frac{351 B_{30}}{152}-\frac{819}{760}\right) A_{11}+\left(\frac{153 A_{03}}{190}+\frac{105 A_{21}}{38}+\frac{216B_{30}}{95}\right.\\
&\quad\left.+\frac{27}{95}\right) B_{02}+\left(-\frac{1287 A_{12}}{760}-\frac{51 B_{21}}{95}-\frac{81 B_{03}}{380}+\frac{2817}{760}\right) B_{11}+\left(-\frac{279 A_{03}}{380}-\frac{18}{19}-\frac{135 B_{30}}{38}-\frac{1143 A_{21}}{380}\right)\\
&\quad B_{20}\Bigg)A_{20}+\Bigg(-\frac{63}{190} \left(\frac{5 A_{12}}{84}+B_{03}+\frac{41 B_{21}}{14}-\frac{197}{28}\right) A_{02}+\left(-\frac{351 A_{03}}{760}+\frac{9 A_{21}}{152}+\frac{9}{76}-\frac{27 B_{30}}{190}\right) A_{11}\\
&\quad +\left(\frac{18 A_{03}}{95}+\frac{33 A_{21}}{38}+\frac{27 B_{30}}{76}+\frac{9}{20}\right) B_{02}+\left(-\frac{81 A_{03}}{190}+\frac{189 B_{30}}{380}+\frac{9}{76}-\frac{21 A_{21}}{95}\right) B_{20}\Bigg) B_{11} \\
&\quad+\Bigg(B_{02}\left.-\frac{3 A_{12}}{95}+\frac{261 B_{03}}{190}+\frac{201 B_{21}}{190}-\frac{36}{95}\right)-\frac{621}{380} A_{02} \left(A_3+\frac{251 A_{21}}{207}+\frac{96 B_{30}}{23}+\frac{2}{3}\right)\Bigg) B_{20}\Bigg)\\
&\quad B_{01}+\left(-\frac{157}{95} B_{02}^3-\frac{191}{190} A_{02}^2 B_{02}\right) B_{20}-\frac{28 B_{02}^4}{95}\Bigg]+648 A_{03}+\Bigg[936 \left(A_{03}+\frac{5 A_{21}}{13}+\frac{3 B_{30}}{13}+\frac{9}{52}\right)\\
&\quad A_{02}+B_{02} \left(-168 A_{12}-72 B_{03}+408 B_{21}+504\right)+B_{20} \left(-120 A_{12}+72 B_{03}+456 B_{21}+648\right)\Bigg] A_{11} \\
& \quad-216 A_{21}+1872 \left(A_{03}+\frac{3 B_{30}}{13}-\frac{5 A_{21}}{13}-\frac{31}{52}\right) A_{02} B_{02}+B_{11}^3 \left(120 B_{02}+120 B_{20}\right)B_{11} \Bigg(480 B_{02}^3\\
&\quad(-432 A_{02}^2 B_{02}+960 B_{20}^2 B_{02}+480 B_{20}^3)+\left(960 B_{02}^2-240 A_{02}^2\right) B_{20}+A_{11}^2 \left(-24 B_{02}-24 B_{20}\right)+A_{11}\\
&\quad\left.-456 A_{02}^2-24 B_{02}^2+456 B_{20}^2+432 B_{02} B_{20}\right)\Bigg)+B_{11}^2 \left(-24 A_{12}+72 B_{03}+120 B_{21}-72\right) +B_{20}^2 \\
&\quad\left.(-240 A_{12}\right.\left.-720 B_{03}+480 B_{21}+1440\right)+B_{02}^2 \left(-384 A_{12}+1440 B_{03}+912 B_{21}+720\right)+A_{11}^2 \\
&\quad\left.(120 A_{12}-72 B_{03}-24 B_{21}+72\right)+B_{20} \Bigg(B_{02} \left(-240 A_{12}+144 B_3+1200 B_{21}+2160\right))\\
&\quad(-288 A_{02} \left(A_{21}+1\right)\Bigg)+A_{02} \left(-720 A_{13}\right.
\left.-432 A_{31}-2880 B_{04}-288 B_{22}\right)+648 B_{30}+B_{21} \\
&\quad\left(216 A_{03}+216 A_{21}+216 B_{30}+270\right)+A_{12} \left(-216 A_{03}-216 A_{21}-216 B_{30}-162\right)+B_{03} \\
&\quad\left(-648 A_{03}\right.\left.+216 A_{21}-648 B_{30}+378\right)+B_{11} \Bigg(648 \left(\frac{19 A_{12}}{27}+B_{03}+\frac{1}{9}-\frac{5 B_{21}}{27}\right) A_{02} \\
& \quad+A_{11} \left(288 A_{03}+288 B_{30}\right)+B_{02} \left(360 A_{03}+72 A_{21}+792 B_{30}+558\right)+B_{20} \left(216 A_3+216 A_{21}\right.\\
&\quad\left.+936 B_{30}+414\right)\Bigg)+B_{01} \Bigg[1224 \left(A_3+\frac{65 A_{21}}{34}+\frac{81 B_{30}}{68}+\frac{319}{204}\right) A_{02}^2+216 \left(-\frac{13 A_{12}}{4}+B_{03}\right.\\
&\quad\left.-\frac{15 B_{21}}{2}-\frac{89}{4}\right) B_{02} A_{02}+\Bigg(7344 \left(A_{03}+\frac{401 A_{21}}{612}+\frac{7}{24}-\frac{29 B_{30}}{136}\right) A_{02}+B_{20} \left(288 A_{12}\right.\\
&\quad \left.+7110 B_{03}\right.\left.+1062 B_{21}-8460\right)+A_{11} \left(414 A_{12}-4185 B_{03}-909 B_{21}+72\right)+B_{02} \left(-1494 A_{12}\right.\\
&\quad \left.+3222 B_{03}\right.\left.-1062 B_{21}-8730\right)+B_{11} \left(2817 A_3+225 A_{21}-414 B_{30}+42\right)\Bigg) A_{20}+\Bigg(2457 \left(A_{03}\right.
\end{align*}
\begin{align*}
&\quad \left.+\frac{83 A_{21}}{819}\right.\left.-\frac{110 B_{30}}{273}+\frac{146}{819}\right) A_{02}+A_{11} \left(243 A_{12}+189 B_{03}+171 B_{21}-603\right)+B_{02} \left(-171 A_{12}\right.\\
&\quad\left.-378 B_{03}\right.\left.-504 B_{21}-99\right)+B_{20} \left(-81 A_{12}+504 B_{03}-1026 B_{21}+1179\right)\Bigg) B_{11}+A_{11}^2 \left(252 A_{03}\right.\\
&\quad \left.-30 A_{21}\right.\left.+477 B_{30}+243\right)+B_{02}^2 \left(-288 A_{03}-360 A_{21}+1188 B_{30}-300\right)+B_{11}^2 \left(315 A_{03}-21 A_{21}\right.\\
&\quad\left.-522 B_{30}-468\right)+A_{20}^2 \left(5004 A_{03}+3468 A_{21}-3816 B_{30}-564\right)+B_{20}^2 \left(-738 A_{03}-1122 A_{21}\right.\\
&\quad\left.-5076 B_{30}-888\right)+A_{11} \Bigg(-2187 \left(\frac{44 A_{12}}{243}+B_{03}+\frac{25 B_{21}}{81}-\frac{14}{81}\right) A_{02}+B_{02} \left(-612 A_{03}\right.\\
&\quad\left.-978 A_{21}+207 B_{30}-321\right)+B_{20} \left(-684 A_{03}-378 A_{21}-1953 B_{30}-933\right)\Bigg)+B_{20} \Bigg(B_{02} \\
&\quad\left(-738 A_3-210 A_{21}-1116 B_{30}-720\right)-1260 A_{02} \left(-\frac{11 A_{12}}{35}+B_{03}+\frac{3 B_{21}}{5}+\frac{23}{7}\right)\Bigg) \\
&\quad +A_{11} \left(-432 A_{04}+720 A_{30}+288 B_{31}-144\right)+B_{20} \left(2880 A_{30}+432 B_{13}+720 B_{31}+288\right) \\
& \quad+B_{02} \left(-864 A_{04}+3168 A_{30}+1296 B_{13}+1008 B_{31}+576\right) +B_{11} \left(-288 A_{13}-720 B_{04}\right.\\
&\quad\left.+144 B_{22}\right.\left.+432 B_{40}\right)+B_{01} \Bigg[-486 A_{03}^2-945 A_{03}+81 A_{12}^2-216 A_{21}^2-1458 B_{03}^2-108 B_{21}^2\\
&\quad-729 B_{30}^2+2673 B_{03}+A_{12} \left(405 B_{03}-432\right)+\left(-135 A_{12}-162 B_{03}+513\right) B_{21}+A_{11} \left(288 A_{13}\right.\\
&\quad\left.+576 A_{31}\right.\left.+576 B_{04}+288 B_{22}\right)+\left(-1053 A_{03}-648\right) B_{30}+A_{21} \left(-1026 A_{03}-1053 B_{30}-405\right)\\
&\quad+B_{11}\left(-1008 A_{04}+720 A_{30}-864 B_{13}-288 B_{31}-432\right)+A_{20} \left(-3168 A_{04}-6048 A_{30}\right.\\
&\quad \left.-1008 B_{13}\right.\left.-720 B_{31}-1728\right)+A_{02} \left(-6336 A_{30}-1008 B_{13}-1296 B_{31}-864\right)+B_{02} \left(288 A_{13}\right.\\
&\quad\left.+864 A_{31}\right.\left.+576 B_{22}+1152 B_{40}\right)+B_{20} \left(-576 A_{31}+1152 B_{04}-576 B_{22}-2304 B_{40}\right)-1728\Bigg]\\
&\quad-54\Bigg].
\end{align*}

\appendix
\renewcommand\thefigure{\thesection.\arabic{figure}}
\setcounter{figure}{0}

\bibliography{MAIN_birhythmicity_isolas}

\end{document}